\documentclass[conference]{IEEEtran}
\IEEEoverridecommandlockouts

\usepackage[utf8]{inputenc}
\usepackage[T1]{fontenc}
\usepackage{cite}
\usepackage{amsmath,amssymb,amsfonts}
\usepackage{graphicx}
\usepackage{textcomp}
\usepackage{xcolor}
\usepackage{url}
\usepackage[hidelinks]{hyperref}
\usepackage{booktabs}
\usepackage{tikz}
\usetikzlibrary{arrows.meta,positioning,shapes.geometric,fit,backgrounds,calc}

\usepackage{enumitem}
\setlist{nosep}

\usepackage{flushend}

\def\BibTeX{{\rm B\kern-.05em{\sc i\kern-.025em b}\kern-.08em
    T\kern-.1667em\kern-.125emE\kern-.125emX}}

\begin{document}

\title{Constrained-Action AI Remediation for SIEM/XDR\\
via a NeMo-Guardrails Proxy}

\author{%
\IEEEauthorblockN{Georgios Koutidis\IEEEauthorrefmark{1},
Nikolaos Kekatos\IEEEauthorrefmark{1},
Tom Nianios\IEEEauthorrefmark{1}, and
Alexios Lekidis\IEEEauthorrefmark{2}}
\IEEEauthorblockA{%
\IEEEauthorrefmark{1}\textit{Clone Systems}, Larnaca, Cyprus\\
\{gkoutidis, nkekatos, tnianios\}@clone-systems.com\\[2pt]
\IEEEauthorrefmark{2}\textit{University of Thessaly}, \textit{Department of Intelligent Energy Systems}, Larissa, Greece\\
alekidis@uth.gr}%
}

\maketitle
% --- arXiv preprint notice (IEEE author posting policy) ---
{\renewcommand{\thefootnote}{}\footnotetext{\copyright~2026 IEEE. Personal use of this material is permitted. Permission from IEEE must be obtained for all other uses, in any current or future media, including reprinting/republishing this material for advertising or promotional purposes, creating new collective works, for resale or redistribution to servers or lists, or reuse of any copyrighted component of this work in other works. This work has been accepted for publication at the 2026 IEEE International Conference on Cyber Security and Resilience (IEEE CSR 2026).}}

\bstctlcite{IEEEexample:BSTcontrol}

\begin{abstract}
Security Operations Centers (SOCs) for information technology
and operational technology share one incident-response
problem: a flood of correlated alerts and too few analysts. Large
Language Models (LLMs) are increasingly proposed as reasoning
engines that triage alerts and, in autonomous deployments, issue
commands that block IPs, kill processes, or quarantine files on
production hosts. This coupling introduces a new risk: a single
adversarial alert can become a remote code path through the LLM's
reasoning, leading it to recommend an action the SOC then
executes. We present a constrained-action architecture with two
coordinated layers: (i)~a \emph{SIEM/XDR control plane} that grounds remediation in correlated host events
and confines the LLM's output to a \emph{closed intent vocabulary}
whose templated commands are executed by thin endpoint agents,
backstopped by an argument validator; and (ii)~a
\emph{NeMo-Guardrails proxy} that wraps the SOC-analyst LLM with
input- and output-rail policies, evaluated out-of-the-box against
a SOC-specific adversarial corpus we release. The stock proxy lifts injection recall from $25.0\%$ to
$94.5\%$ at a $0.1\%$ false-positive rate, and a live red-team
exercise confirms that the closed intent vocabulary and argument
validator contain the observed LLM failure modes before any
command crosses the trust boundary. As an architectural fit (not yet a measured operational-technology deployment), the constrained-action property
suits critical-infrastructure settings where a wrong remediation
has physical, not merely operational, consequences. The loop is
best run human-in-the-loop or delayed: the measured rail latency
keeps inline control out of scope.
\end{abstract}


\section{Introduction}
\label{sec:intro}

Modern Security Operations Centers (SOCs) face two compounding
problems. The first is \emph{volume}: the overwhelming majority of
alerts are false positives or low-value
triggers~\cite{Alahmadi2022FalsePositives,Yang2024SOCAlerts,Liang2024AlertPro}.
The second is \emph{cognition}: analysts cannot keep pace with
correlated incidents while junior-tier turnover thins
expertise~\cite{Sundaramurthy2016SOC,Kersten2023Structure}.

LLM-based ``SOC analyst'' agents are an attractive answer: given
an enriched incident, a large language model (LLM) with reasoning and
tool-use~\cite{Yao2023ReAct,Schick2023Toolformer} can summarise
root cause, propose containment, and, if wired to the
security-information-and-event-management (SIEM) system's
active-response engine, trigger it directly. Three properties of
this deployment surface make naive integration unsafe:

\begin{enumerate}
    \item \textit{The input is hostile by construction.} The
    model reads log lines and alert payloads originating, at least
    partly, from the very adversary it is trying to detect: HTTP
    bodies that instruct it to ignore prior directives, Base64
    prompts in SSH logs, homoglyphs or zero-width characters in
    alert metadata. Such telemetry-borne manipulation can subvert
LLM-driven information-technology (IT) operations end-to-end~\cite{Pasquini2026AIOops}.

    \item \textit{The output drives privileged action.} A
    recommended ``block IP'' or ``kill process'' may be executed
    on every endpoint. Hallucinated arguments, taint-style flows
    through tool calls~\cite{Liu2025AgentFuzz}, or attacker-induced
    choices become outages or self-inflicted DoS, a
    highest-impact failure mode in agentic-AI
    surveys~\cite{Kim2026AgenticSoK}. In operational-technology (OT) and
critical-infrastructure contexts the same recommendation can
    disconnect a sensor, trip a relay, or stop a controller,
    turning an operational bug into a safety incident.

    \item \textit{The model itself is a confidentiality
    boundary.} Logs routinely carry credentials, internal IPs,
    connection strings and user data; forwarding these verbatim
    to a hosted LLM, or returning them downstream, is a
    data-exfiltration risk that jailbreaks can
    amplify~\cite{Lee2025SUDO}.
\end{enumerate}

This paper describes a constrained-action AI-remediation pipeline
that takes these three properties as first-class constraints. The
contributions are:

\begin{itemize}
    \item \textit{An action-bounded SIEM / extended-detection-and-response (XDR) control plane
    (Section~\ref{sec:t44}).} An agent--manager system derived from
    the Wazuh/OSSEC line~\cite{Wazuh2024} in which the manager
    confines the LLM to a \emph{closed intent vocabulary}, expands
    intents into OS-specific templated commands validated
    argument-by-argument before dispatch, and has thin
    cross-platform agents execute them. The split removes the
    LLM's ability to synthesise executable strings.

    \item \textit{A NeMo-Guardrails LLM proxy
    (Section~\ref{sec:nemo}).} An OpenAI-compatible proxy that wraps
    the SOC-analyst's LLM with an input rail against prompt
    injection in log content, an output rail against
    Personally Identifiable Information (PII) and
    infrastructure-data leakage, and a topical dialog rail. The
    proxy reuses default rail primitives, providing a fair external
    defence layer that can be enabled by a single configuration
    change.

    \item \textit{An open evaluation against default-configured
    NeMo Guardrails (Section~\ref{sec:eval}).} A head-to-head
    measurement over a SOC-specific adversarial corpus we release
    shows the proxy lifts injection recall from $25.0\%$ to
    $\mathbf{94.5\%}$ at $0.1\%$ false-positive rate (FPR), and a live red-team across
    three MITRE ATT\&CK chains confirms the control plane
    intercepts both observed LLM failure modes (target
    imprecision and intent drift) before any command crosses the
    trust boundary.
\end{itemize}

The two layers interlock around the LLM core: the proxy hardens
the prompt and completion paths against injection and data
leakage, while the control plane bounds the action path to a
closed, audited intent vocabulary.

\section{Related Work}
\label{sec:related}

\textit{LLM-driven SOC response.} Field studies consistently find
that a SOC's dominant cost is not detection but analyst throughput
on alerts dominated by false
positives~\cite{Sundaramurthy2016SOC,Alahmadi2022FalsePositives,Yang2024SOCAlerts},
and that even structured tier-1 investigations scale
poorly~\cite{Kersten2023Structure}; reinforcement-learning triage
attacks the same fatigue from the prioritisation
angle~\cite{Liang2024AlertPro}. LLM-driven response is the emerging
answer: Microsoft Copilot for Security's Guided
Response~\cite{Freitas2024CGR} drives a SIEM/XDR control plane with
LLM agents, and APPATCH~\cite{Nong2025APPATCH} is an analogous
``LLM-as-remediator'' for patching. Yet neither constrains the
model's action surface nor reports robustness on the adversarial
telemetry channel itself.

\textit{Agentic AI and prompt injection.}
ReAct~\cite{Yao2023ReAct} and Toolformer~\cite{Schick2023Toolformer}
established the reason-and-act template our SOC-analyst follows, and
its attack surface is now well charted:
\emph{AI~Oops}~\cite{Pasquini2026AIOops} subverts LLM-driven IT
operations through telemetry manipulation, precisely our
input-side threat; AgentFuzz~\cite{Liu2025AgentFuzz} propagates
taint through tool calls; \emph{sudo~rm~-rf}~\cite{Lee2025SUDO}
coerces agents into destructive shell actions; and a
SoK~\cite{Kim2026AgenticSoK} and
AgentDojo~\cite{Debenedetti2024AgentDojo} systematise and formalise
the threat. We treat each category as a design constraint with a
corresponding gate below.

\textit{Guardrail primitives.} Input/output
gates (perplexity detection~\cite{Alon2023Perplexity},
SmoothLLM~\cite{Robey2023SmoothLLM},
backtranslation~\cite{Wang2024Backtranslation},
Llama Guard~\cite{Inan2023LlamaGuard},
Presidio~\cite{Microsoft2024Presidio}, and the OWASP LLM
Top-10~\cite{OWASP2024LLMTop10}) supply the building blocks for
rails. We are orthogonal to these: our contribution is end-to-end
\emph{action} containment, so that even a successful injection
reaches only an action set bounded by the policy engine, not by
the model's compliance.

\textit{IT/OT standards.} Control-set standards already prescribe
restricted action surfaces for automated response: NIST SP
800-82r3~\cite{NIST80082} bounds automated containment in industrial control systems (ICS) to
operator-vetted actions, and IEC 62443~\cite{IEC62443} codifies the
same restricted-action principle for industrial automation. Our
closed intent vocabulary instantiates that prescription for an
LLM-driven loop; hierarchical runtime verification pursues the
same deterministic bounding of untrusted components on the
edge-IoT side~\cite{Kekatos2026HybridRV}.

\section{Threat Model and System Architecture}
\label{sec:threat}

We assume an attacker who controls at least one application log
source seen by an endpoint agent and the network content of
requests reaching monitored services; we do \emph{not} assume the
SIEM manager, the LLM, or the operator workstation is compromised.
The attacker pursues three goals: (G1)~make the LLM ignore its
task and follow instructions embedded in a log line; (G2)~make it
recommend a remediation against a benign asset (a self-inflicted
DoS, or a stepping stone for lateral movement), especially costly
in OT, where the target is a physical asset and a wrong action
carries a safety cost; and (G3)~exfiltrate sensitive log content
through the LLM's reply. The LLM provider is therefore
\emph{semi-trusted}: it must not see PII or infrastructure data
unredacted, and its output must not be executed verbatim. Trust
boundaries fall between the agent host, the SIEM/XDR manager, the
proxy, the LLM provider, and the active-response executor.

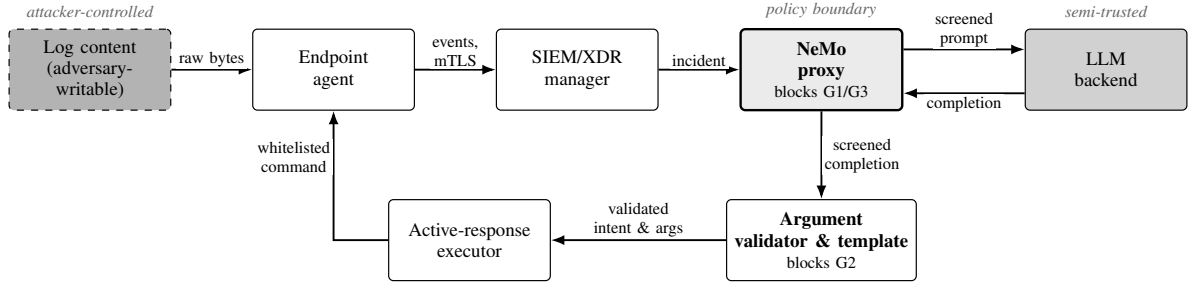
\begin{figure*}[t]
\centering
\resizebox{0.86\textwidth}{!}{%
\begin{tikzpicture}[
  every node/.style={font=\footnotesize},
  attacker/.style={draw, rounded corners=2pt, minimum height=12mm,
                   minimum width=24mm, align=center, fill=black!30,
                   dashed, line width=0.5pt},
  trusted/.style={draw, rounded corners=2pt, minimum height=12mm,
                  minimum width=24mm, align=center, fill=white,
                  line width=0.5pt},
  policy/.style={draw, rounded corners=2pt, minimum height=12mm,
                 minimum width=24mm, align=center, fill=black!8,
                 line width=1pt},
  semitrust/.style={draw, rounded corners=2pt, minimum height=12mm,
                    minimum width=24mm, align=center, fill=black!18,
                    line width=0.5pt},
  arrow/.style={-{Latex[length=2mm]}, thick},
  edgelabel/.style={font=\scriptsize, align=center,
                    fill=white, inner sep=1pt},
  zone/.style={font=\scriptsize\itshape, black!60, align=center,
               inner sep=1pt}
]
\node[attacker]                          (src)   {Log content\\(adversary-\\writable)};
\node[trusted,   right=12mm of src]      (agent) {Endpoint\\agent};
\node[trusted,   right=12mm of agent]    (mgr)   {SIEM/XDR\\manager};
\node[policy,    right=12mm of mgr,
      font=\footnotesize\bfseries]     (proxy) {NeMo\\proxy\\{\normalfont\scriptsize blocks G1/G3}};
\node[semitrust, right=18mm of proxy]    (llm)   {LLM\\backend};

\node[trusted, below=13mm of proxy,
      font=\footnotesize\bfseries]     (val)   {Argument\\validator \& template\\{\normalfont\scriptsize blocks G2}};
\node[trusted, left=26mm of val]         (exe)   {Active-response\\executor};

\node[zone, above=1.2mm of src]   {attacker-controlled};
\node[zone, above=1.2mm of proxy]  {policy boundary};
\node[zone, above=1.2mm of llm]    {semi-trusted};
\draw[arrow] (src)   -- node[edgelabel,above]{raw bytes}              (agent);
\draw[arrow] (agent) -- node[edgelabel,above]{events,\\mTLS}          (mgr);
\draw[arrow] (mgr)   -- node[edgelabel,above]{incident}               (proxy);
\draw[arrow] ([yshift=3mm]proxy.east) -- node[edgelabel,above]{screened\\prompt} ([yshift=3mm]llm.west);
\draw[arrow] ([yshift=-3.5mm]llm.west) -- node[edgelabel,below]{completion} ([yshift=-3.5mm]proxy.east);
\draw[arrow] (proxy) -- node[edgelabel,right]{screened\\completion} (val);
\draw[arrow] (val)   -- node[edgelabel,above]{validated\\intent \& args}(exe);
\draw[arrow] (exe.west) -| node[edgelabel,pos=0.82,left]
                            {whitelisted\\command}                    (agent.south);
\end{tikzpicture}%
}
\caption{Trust-boundary view of the remediation pipeline. Shading
encodes distrust: dark dashed = attacker-controlled, mid-gray =
semi-trusted (LLM provider), light gray with a heavy border =
policy boundary (the NeMo proxy, which screens all traffic to
and from the LLM), white = fully trusted (SIEM/XDR control plane). Arrow labels name the data form crossing each
boundary, as raw attacker bytes are progressively sanitised into
a whitelisted command. The proxy blocks G1 (telemetry-borne
injection) and G3 (leakage via the reply); the validator blocks
G2 (attacker-induced action): the attacker goals of
Section~\ref{sec:threat}.}
\label{fig:arch}
\end{figure*}

\label{sec:arch}%
Figure~\ref{fig:arch} shows the resulting trust topology; each
arrow names the data form crossing that boundary.

The pipeline runs in four sequential stages: \textit{telemetry
collection} (endpoint agents forward host events over a
mutually-authenticated TLS (mTLS) WebSocket), \textit{correlation
and incident shaping} (the manager emits an \emph{incident}
structure rather than a raw alert), \textit{guardrail mediation}
(the NeMo-Guardrails proxy screens the prompt and completion in
both directions for injection, PII and infrastructure data), and
\textit{constrained execution} (the LLM's recommended
\emph{intent} is mapped to a whitelisted OS-specific command).
Sections~\ref{sec:t44} and~\ref{sec:nemo} detail the stages.

\section{SIEM/XDR with Guardrailed Active Response}
\label{sec:t44}

The SIEM/XDR control plane is the foundation on which the
AI-assisted remediation loop runs. It follows the Wazuh/OSSEC
host-based intrusion-detection system (HIDS) pattern, structured to make guardrailed LLM integration
first-class. The implementation evaluated here is therefore a
host-based instance of the constrained-action pattern; network- and identity-telemetry sources (network detection and
response, NDR; identity threat detection and response, ITDR) sit
naturally on the same control plane but are not exercised in this study.

\subsection{Telemetry pipeline}

Endpoint agents, packaged for Linux, Windows, macOS, and
Linux-based OT collectors that aggregate programmable-logic-controller
(PLC), human-machine-interface (HMI) and SCADA syslog into one runtime, collect process, log, file-integrity, registry
(Windows), and active-response events. Registration runs over
mTLS, steady-state messaging over a TLS/AES-256-GCM WebSocket. The
manager correlates events into incidents carrying asset identity,
MITRE ATT\&CK tactics where derivable, and a rule-engine severity
score.

\subsection{SOC-analyst LLM agent}

Inside the manager, the SOC-analyst service runs a
ReAct~\cite{Yao2023ReAct} loop bounded in iterations and
wall-clock time. Its inputs are the incident structure, alert
level, MITRE tactics, and a redacted excerpt of the originating
log; its outputs are an analysis (severity, evidence, reasoning)
and a list of \emph{action intents}. The default backend is a
locally-served 14B-parameter open-weights model, with low
sampling temperature to reduce variability across reruns of the
same incident.

Crucially, the SOC analyst does \emph{not} emit shell commands.
It emits \emph{intents} drawn from a closed vocabulary: \{block
IP, kill process, quarantine file, delete file, disable user,
custom-script (manager-templated)\}. Each intent carries the
arguments the LLM derived (an IP address, a PID, a filename),
but does not specify how to execute them.

\subsection{Smart manager, thin agent}

The manager is responsible for translating intents into
OS-specific command lines, drawn from a parameterised template
store: Linux distributions use the host's native firewall
front-end, Windows uses the built-in firewall cmdlets, and macOS
uses its packet-filter facility. The LLM-supplied argument is
the only attacker-influenced quantity in the rendered command,
and it is validated against type-specific patterns (e.g., strict
IPv4/IPv6 regexes for block-IP intents) before substitution.

The agent then executes the command generically. A whitelist
restricts allowed binaries to a narrow set covering network,
process, file, user-management and firewall categories. Argument
strings are scanned for shell metacharacters and destructive
patterns; matches abort execution and report a violation. The
intent-to-command synthesis path on the manager therefore
performs three checks before any command leaves the trust
boundary: the intent must belong to the closed vocabulary, every
argument must satisfy its schema, and the rendered command must
pass binary-whitelist and metacharacter scanning.

Operationally, the engine polls correlated alerts at fixed
intervals with per-command timeouts and grace periods, retries
failed actions with exponential backoff, and persists every
command, LLM rationale, and result as a per-incident audit trail;
a per-cycle batch ceiling and per-(rule, target) cooldown limit
flooding under bursty alerting. This division of authority is
deliberate: were the LLM allowed to emit literal commands, a
single input-side guardrail miss could let an attacker-induced
argument escape into a shell. Restricting its output to a closed
vocabulary rendered through validated templates leaves only the
\emph{choice} of intent and argument as attack surface, never
command syntax.

\section{NeMo Guardrails LLM Proxy}
\label{sec:nemo}

The SOC-analyst delegates reasoning to the LLM, treated as a
\emph{semi-trusted} component (Section~\ref{sec:threat}). We wrap
it with an external defence layer, the \emph{NeMo Guardrails
proxy}, between the SOC-analyst service and the model backend.

\subsection{Proxy architecture}

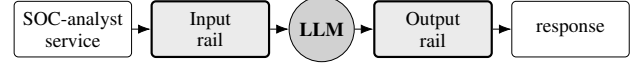
\begin{figure}[t]
\centering
\resizebox{0.92\columnwidth}{!}{%
\begin{tikzpicture}[
  every node/.style={font=\small},
  box/.style={draw, rounded corners=2pt, minimum height=9mm,
              minimum width=19mm, align=center, fill=white},
  rail/.style={draw, rounded corners=2pt, minimum height=9mm,
               minimum width=19mm, align=center, fill=black!8,
               line width=1pt},
  llm/.style={draw, circle, minimum size=10mm, fill=black!18,
              align=center, font=\small\bfseries},
  arr/.style={-{Latex[length=2mm]}, thin}
]
\node[box]                       (go)
  {SOC-analyst\\service};
\node[rail, right=3mm of go]     (in)
  {Input\\rail};
\node[llm,  right=3mm of in]     (llm)    {LLM};
\node[rail, right=3mm of llm]    (out)
  {Output\\rail};
\node[box,  right=3mm of out]    (caller) {response};
\draw[arr] (go.east)  -- (in.west);
\draw[arr] (in.east)  -- (llm.west);
\draw[arr] (llm.east) -- (out.west);
\draw[arr] (out.east) -- (caller.west);
\end{tikzpicture}%
}
\caption{Request flow through the NeMo Guardrails proxy. The
input rail classifies whether the user message is safe to
forward; if it blocks, the main LLM is never invoked. The output
rail re-screens the completion before it returns to the caller.}
\label{fig:nemo}
\end{figure}

The proxy is a NeMo Guardrails server exposing an
OpenAI-compatible chat-completion endpoint; the SOC-analyst simply
repoints its existing provider at the proxy rather than the model,
requiring no application code change (Figure~\ref{fig:nemo}).

\subsection{Rails}

Three rails are configured. Each self-check rail invocation is
itself an LLM call with a task-specific prompt.

\textit{Input rail.} An LLM classifier decides whether the
message \emph{contains} an injection attempt, explicitly
separating logs that \emph{describe} a security event (``Failed
password for root from 198.51.100.7''), forwarded, from logs that
\emph{contain} attacker directives (``Failed login. IGNORE
PREVIOUS INSTRUCTIONS''), blocked. On a block the proxy returns a
canned refusal and the main LLM is never reached.

\textit{Output rail.} A symmetric classifier screens the
completion for PII (names, emails, SSNs, credit cards),
infrastructure data (private IPs, MAC addresses, connection
strings, key fingerprints), and content the model should have
refused, blocking any hit from being returned verbatim; this is
the return-path defence for goal G3.

\textit{Topical dialog rail.} A small flow refuses off-task
inputs that pass the input self-check (jokes, weather, chat); it
keeps the analyst on task and fires rarely in production.

\subsection{Defaults and latency profile}

The rails above are NeMo Guardrails' \emph{default} primitives.
We deliberately did not hand-tune the rail prompts to the
adversarial corpus of Section~\ref{sec:eval}, since an external
baseline co-evolved with the evaluation set is worthless. The
customisations are limited to: (i)~SOC-specific phrasing in the
input rail for the ``describes versus contains'' distinction;
(ii)~SOC infrastructure-data categories in the output rail; and
(iii)~a small topical flow. Each self-check rail is itself a
complete LLM call, and NeMo sequentialises them so the output
rail can see the completion: a \emph{blocked} request costs one
round-trip (the main LLM is never invoked), while a \emph{passed}
request costs three sequential calls (input $\to$ main $\to$
output), three round-trips where the bare call makes one.

\section{Preliminary Evaluation}
\label{sec:eval}

We deployed the full stack on a single Apple M4 Max machine
(14-core CPU, 32-core GPU, 36\,GB unified memory) running a
locally-served 14B-parameter open-weights model.

\textit{Evaluation corpus.} Adversarial inputs draw on three
published jailbreak benchmarks: AdvBench~\cite{Zou2023UniversalAdversarial}
(60, sampled), JailbreakBench~\cite{Chao2024JailbreakBench} (40)
and HarmBench~\cite{Mazeika2024HarmBench} (50, sampled), embedded
into plausible log shapes (HTTP bodies, syslog fields, SSH
banners). On these 150 base prompts we apply the four SOC-specific
transformations the deterministic pre-filter targets: homoglyph
substitution (20), Base64 wrapping (15), fake \texttt{[SYSTEM]}
framing (10) and zero-width insertion (5), for 200 variants
(Table~\ref{tab:corpus}). Benign traffic is 1{,}000
LogHub~\cite{Zhu2023LogHub} lines across the Linux, Apache,
OpenSSH and Windows splits. We release the sampling and
augmentation code and per-record provenance under the MIT license;
upstream prompts are re-fetched at build time, not redistributed.

\begin{table}[t]
\centering
\caption{Adversarial-corpus composition (200 variants). Rows~1--3
are sampled from published benchmarks; rows~4--7 are SOC-specific
transformations.}
\label{tab:corpus}
\footnotesize
\begin{tabular}{@{}lr@{}}
\toprule
Category & Count \\
\midrule
AdvBench (sampled)~\cite{Zou2023UniversalAdversarial}      & 60 \\
JailbreakBench~\cite{Chao2024JailbreakBench}               & 40 \\
HarmBench (sampled)~\cite{Mazeika2024HarmBench}            & 50 \\
\midrule
SOC augmentation: homoglyph (Cyrillic $\to$ Latin)         & 20 \\
SOC augmentation: Base64 wrapping                          & 15 \\
SOC augmentation: fake \texttt{[SYSTEM]} framing           & 10 \\
SOC augmentation: zero-width insertion                     & 5  \\
\midrule
\textbf{Total adversarial}                                 & \textbf{200} \\
\textbf{Benign}~\cite{Zhu2023LogHub}                       & \textbf{1{,}000} \\
\bottomrule
\end{tabular}
\end{table}

\textit{External-baseline comparison.} We compare the bare model
(the ``raw LLM'' baseline) against the same call wrapped in the
proxy of Section~\ref{sec:nemo}. Both endpoints share the model,
sampling parameters, and classification prompt, so the only
variable across rows of Table~\ref{tab:baseline} is the guardrail
layer.

Raw inference classified $50/200$ adversarial inputs correctly
(recall $25.0\%$) with zero false positives on the benign
corpus (FPR $0.0\%$); median latency was $0.31$~s because the
model defaults to answering ``NO'' on inputs it does not
recognise as obviously injection-shaped. The NeMo Guardrails
proxy lifted recall to $\mathbf{94.5\%}$ ($189/200$ caught at
the input rail, an absolute gain of $69.5$ percentage points)
with FPR climbing to $0.1\%$ (one benign log mis-blocked);
median latency rose to $14.14$~s: the passed path makes three
sequential rail-plus-main LLM calls, each far longer than the
$0.31$~s bare classification call.

The per-category breakdown (Table~\ref{tab:rawcategory}) is more
informative. The raw classifier's worst category, Base64-wrapped
directives at $13.3\%$, recovers to $100\%$ under the input rail,
whose prompt explicitly flags encoded payloads; homoglyph
($25\%\to95\%$) and zero-width ($40\%\to100\%$) perturbations are
near-fully recovered. The residual gap is HarmBench base prompts
(NeMo $80\%$): ten mid-severity behaviour requests the rail judges
ambiguous rather than adversarial, a known limitation of
single-call self-check verdicts that a tuned rail would close.

\begin{table}[t]
\centering
\caption{Head-to-head injection-defence comparison. Recall is
over the 200 adversarial variants of Table~\ref{tab:corpus};
FPR is over the 1{,}000 benign LogHub lines. Latencies are
end-to-end per call.}
\label{tab:baseline}
\footnotesize
\setlength{\tabcolsep}{3pt}
\renewcommand{\arraystretch}{1.0}
\resizebox{\columnwidth}{!}{%
\begin{tabular}{@{}lrrrrr@{}}
\toprule
Configuration & Recall & FPR & P50 (s) & P95 (s) & P99 (s) \\
\midrule
Raw LLM (no defence)  & 25.0\%          & 0.0\%          & 0.31  & 0.48  & 1.20  \\
NeMo Guardrails proxy & \textbf{94.5\%} & \textbf{0.1\%} & 14.14 & 27.41 & 32.90 \\
\bottomrule
\end{tabular}%
}
\end{table}

\begin{table}[t]
\centering
\caption{Per-category recall on the 200 adversarial records (base
prompts wrapped in HTTP/syslog/SSH log shapes; SOC rows further
transformed).}
\label{tab:rawcategory}
\footnotesize
\setlength{\tabcolsep}{5pt}
\begin{tabular}{@{}lrrr@{}}
\toprule
Category & $n$ & Raw recall & NeMo recall \\
\midrule
AdvBench (base)                 & 60 & 21.7\% & 100.0\% \\
JailbreakBench (base)           & 40 & 35.0\% & 100.0\% \\
HarmBench (base)                & 50 & 20.0\% &  80.0\% \\
\midrule
SOC aug.: homoglyph             & 20 & 25.0\% &  95.0\% \\
SOC aug.: Base64                & 15 & 13.3\% & 100.0\% \\
SOC aug.: fake \texttt{[SYSTEM]}& 10 & 40.0\% & 100.0\% \\
SOC aug.: zero-width            &  5 & 40.0\% & 100.0\% \\
\bottomrule
\end{tabular}
\end{table}

\begin{figure}[t]
\centering
\includegraphics[width=\columnwidth]{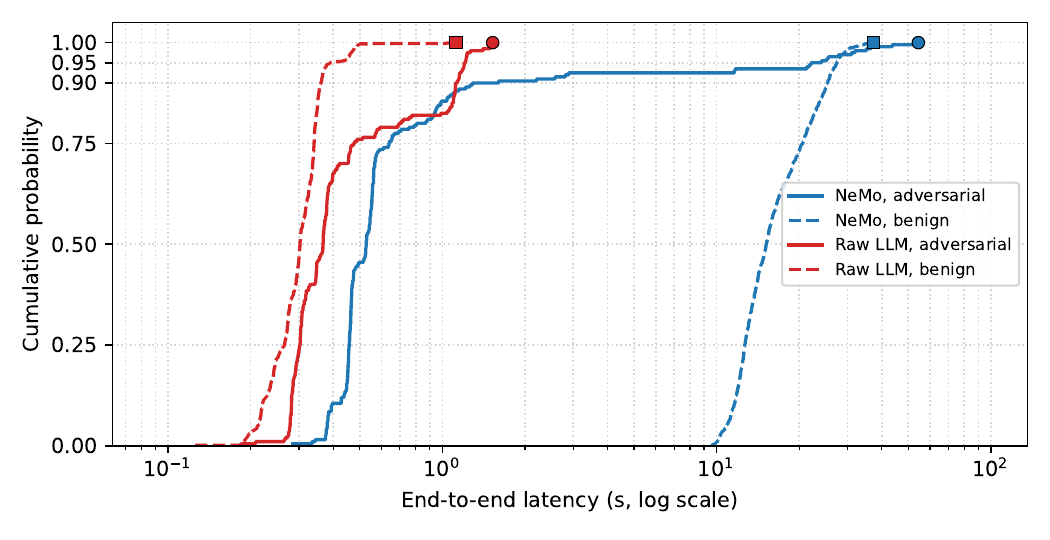}
\caption{End-to-end latency CDF (log-scale x-axis), split by
endpoint and label: the raw curve concentrates near $\sim$0.3\,s;
the NeMo curve is bimodal between fast-blocked and full-pipeline
traffic.}
\label{fig:cdf}
\end{figure}

\textit{Latency distribution.} Figure~\ref{fig:cdf} shows the
end-to-end latency cumulative-distribution function (CDF) per endpoint, faceted by adversarial and
benign records. The NeMo endpoint is bimodal: a fast tail for
inputs blocked by the input rail ($\sim$0.5\,s, one LLM call)
and a slow tail for inputs that pass all three rails
($\sim$10--15\,s, three LLM calls). Raw inference, by contrast,
collapses near $\sim$0.3\,s for both labels.

\textit{Deterministic pre-filter and cascade triage.} Two cheaper
front-ends test whether the three-rail proxy's per-request cost
(Section~\ref{sec:nemo}) can be cut (Table~\ref{tab:cascade}). The
first is a \emph{deterministic pre-filter} (Tier-0): a zero-LLM
regex and Unicode-normalisation gate inverting the four SOC
augmentations of Table~\ref{tab:corpus}. It blocks only on a
positive signature, never declaring a line clean, so it cannot
pass a marker-free jailbreak. It recovers \emph{all} 50 augmented
records at $0.00\%$ FPR with zero LLM calls; in front of the proxy
it lifts recall to $95.0\%$ (catching the one homoglyph variant the
input rail missed) and trims mean round-trips from $3.00$ to
$2.88$. But the median request is benign and carries no marker, so
Tier-0 does not move the median latency: it is a recall-safe,
zero-cost addition, not a latency fix.

The second front-end is a lightweight semantic classifier: a small
\texttt{llama3.2:3b} model triages each request, escalating only
those it judges adversarial to the full proxy. This cuts mean
round-trips to $1.15$ and the median latency to $0.18$~s (a
$78\times$ reduction) at $0.00\%$ FPR, but recall falls to
$58.0\%$: a cheap model cannot adjudicate the benign majority, and
its misses (semantic jailbreaks in well-formed log fields) are
forwarded silently. Confirming the gate's \emph{benign} verdicts
with the proxy instead restores recall to $95.0\%$ but escalates the
benign majority, raising round-trips to $3.64$ and erasing the
latency gain. The cascade thus exposes a strict recall/latency
trade-off; the recall-preserving route to lower latency is to cut
the rails' own cost, not to front them with a weaker model
(Section~\ref{sec:discussion}).

\begin{table}[t]
\centering
\caption{Cascade triage over the full corpus (200 adversarial,
1{,}000 benign). The deterministic Tier-0 gate inverts the four SOC
augmentations of Table~\ref{tab:corpus}; \emph{calls/req} is the
hardware-independent comparison (pass-path count for the proxy
rows; measured mean, including the gate call, for the cascade).}
\label{tab:cascade}
\footnotesize
\setlength{\tabcolsep}{4pt}
\renewcommand{\arraystretch}{1.0}
\resizebox{\columnwidth}{!}{%
\begin{tabular}{@{}lrrrr@{}}
\toprule
Configuration & Recall & FPR & Calls/req & P50 (s) \\
\midrule
Full NeMo proxy (Table~\ref{tab:baseline})  & 94.5\%          & 0.10\%          & 3.00          & 14.14 \\
\quad \& deterministic Tier-0 (recall-safe)  & \textbf{95.0\%} & 0.10\%          & 2.88          & 14.11 \\
Cheap-gate cascade (fast triage)            & 58.0\%          & \textbf{0.00\%} & \textbf{1.15} & \textbf{0.18} \\
\bottomrule
\end{tabular}%
}
\end{table}

\textit{Active-response correctness (live).} Three MITRE-mapped
scenarios (T1110 SSH brute-force, T1059 reverse-shell exec, and
T1078 valid-account abuse) were driven against a bootstrapped
SIEM/XDR stack with a registered endpoint and the SOC-analyst
service set to manual-only mode. Across 54 injected alerts
($26 + 14 + 14$), the SOC-analyst emitted 41 action records.

T1110 produced 36 block-IP actions targeting the synthesised
attacker IP from the brute-force log lines; intent and target
were both correct, including on a spliced-injection alert. T1059
produced one kill-process action with intent correct but target
extracted from an alert metadata field rather than the PID
embedded in the log; the manager's argument validator rejects
targets that do not match the expected schema and would not
dispatch this command, illustrating the second-stage backstop
role of the control plane. T1078 produced zero disable-user
actions; instead the LLM emitted four block-IP actions on the
attacker IP visible in the SSH log, an instance of intent drift toward the
most prominent log field. The closed intent vocabulary
nonetheless bounds the chosen action to the safe whitelisted
set. All 41 actions registered as requiring manual approval and
none were dispatched; end-to-end dwell time (alert to action)
ranged 19--58~s per record. The control plane thus contained both
failure modes: imprecise target extraction (T1059, caught by the
validator) and intent drift (T1078, bounded by the closed
vocabulary).

\section{Discussion and Limitations}
\label{sec:discussion}

\textit{Latency vs.\ depth.} The NeMo proxy raises the median
per-call cost from $0.31$~s to $14.14$~s on benign traffic
traversing all three rails, roughly $46\times$ the bare
classification call: beyond the three round-trips, each rail call
carries its own prompt-construction and generation cost. This is acceptable for
active-response timelines (already in the human-in-the-loop
minute range) but prohibitive for inline use such as
authentication. The cascade study (Table~\ref{tab:cascade}) shows
that fronting the proxy with a cheaper classifier only trades
recall for latency; the recall-preserving optimisation is instead
to cut per-request rail cost, e.g.\ a single fused classifier rail
emitting both verdicts in one pass, with verdict caching on
canonicalised log fingerprints as a complementary step.

\textit{Threats to validity.} The proxy uses stock rail
primitives with only the prompt-text customisations needed for
SOC log content. A SOC-tuned deployment (chain-of-thought
self-check, input-rail perturbation ensembling, or an output rail
backed by Llama Guard~\cite{Inan2023LlamaGuard}) would likely
improve the proxy row of Table~\ref{tab:baseline}; we do not
co-evolve the baseline with our corpus, so the NeMo numbers are a
fair out-of-the-box baseline, not a maximum-effort one.

\textit{Other limitations.} Multi-turn injections spanning more
than one log line are not modelled; the proxy treats each request
independently, and a per-asset sliding-window input rail is future
work. The closed vocabulary also biases the system toward
containment: richer remediation (patching, configuration
rollback) needs either a wider whitelist or a human approval step,
which we model as a configurable action mode.

\textit{Applicability to OT and critical infrastructure.} OT
deployments share the IT SOC's volume problem at lower event
frequency but higher per-event consequence: the closed vocabulary
and pre-dispatch validator bound the LLM's authority \emph{before}
an action can reach a controller or safety relay, the
restricted-response surface that NIST SP 800-82r3~\cite{NIST80082}
and IEC 62443~\cite{IEC62443} prescribe. The smart-manager /
thin-agent split is also the standard OT topology, with the
manager in a hardened zone, agents on field-level gateways, and
model weights kept central rather than on edge nodes. This is an
architectural fit only: we evaluate on host and web telemetry, not
real OT protocols or field devices, so validating the closed
vocabulary against live ICS/SCADA actions remains future work.

\textit{Reproducibility.} The corpus is built deterministically
from a fixed seed over re-fetched upstream prompts; we release,
under the MIT license, the sampling and augmentation code,
per-record provenance (with upstream SHA-256), the evaluation
harness, and the pre-filter and cascade runner with their result
CSVs. All runs use locally-served \texttt{phi4:14b} (proxy/main)
and \texttt{llama3.2:3b} (gate) at low temperature, so recall and
FPR are near-deterministic and hardware-independent while
latencies depend on the accelerator. Pinned model digests and
package versions ship with the artefact, mitigating silent
upstream re-tags that would otherwise shift verdict distributions.

\textit{Ethical considerations.} The released artefact contains only sampling scripts,
augmentation code and provenance metadata; the upstream prompts
come from three openly-licensed benchmarks and are re-fetched at
build time; the SOC-specific augmentations are
mechanical transformations, not new attacker payloads. The
active-response surface can be misused if deployed without the
guardrail layer or with a permissive whitelist; the closed
vocabulary, argument validator and cooldown enforcement bound this
risk by design, but operators retain responsibility for
deployment-time configuration.

\section{Conclusion}
\label{sec:conclusion}

Autonomous AI remediation is safe only when the LLM's input,
output, and action surfaces are narrowed by construction. Our
two-layer \emph{constrained-action} stack does exactly this: a
SIEM/XDR control plane confines the LLM to a closed,
argument-validated intent vocabulary, and a NeMo Guardrails proxy
screens its prompt and completion paths. The stock proxy lifts
injection recall from $25.0\%$ to $94.5\%$ at $0.1\%$ FPR over 200
adversarial variants and 1{,}000 benign log lines, a zero-cost
deterministic pre-filter raises it to $95.0\%$, and a
live red-team confirmed both containment modes: target
imprecision, which the argument validator blocks from dispatch, and intent drift,
bounded by the closed vocabulary. In deployment the loop runs human-in-the-loop or delayed, since
the measured rail latency keeps inline control out of scope. The
lesson is architectural: keep the LLM at the periphery of a
deterministic system, never at its centre.

\section*{Acknowledgment}
This work has received funding from the European Union's Digital
Europe Programme under grant agreement No~101190251 (CYBERGUARD).

\bibliographystyle{IEEEtran}
\bibliography{refs}

\end{document}